\documentclass[%
 reprint,
 amsmath,amssymb,
 aps,
]{revtex4-2}

\usepackage{graphicx}
\usepackage{dcolumn}
\usepackage{bm}
\usepackage{caption}
\usepackage{subcaption}
\usepackage[modulo]{lineno}
\usepackage{placeins}
\usepackage{textcase}
\usepackage{mathtools}

\begin{document}

\preprint{APS/123-QED}

\title{Femtosecond-Scale MeV-UED Beamline\\ Using a Stand-Alone Multi-Cell RF Photogun}

\author{Thomas Geoffrey Lucas$^1$}
 \thanks{Corresponding Author: \\ thomas.lucas@psi.ch}
 \author{Paolo Craievich$^1$}
 \author{David Alesini$^2$}
 \author{Carl Beard$^1$}
 \author{Hans-H. Braun$^1$}
 \author{Alexander Dietrich$^1$}
 \author{Zheqiao Geng$^1$}
 \author{Rasmus Ischebeck$^1$}
 \author{Eduard Prat$^1$}
 \author{Mike Seidel$^{1,3}$}
 \author{Cezary Sydlo$^1$}
 \author{Alexandre Trisorio$^1$}
 \author{Carlo Vicario$^1$}
 \author{Riccardo Zennaro$^1$}
 \affiliation{%
 $^1$ Paul Scherrer Institut, Forschungsstrasse 111 5232Villigen, Switzerland.\\
 $^2$Laboratori Nazionali di Frascati, INFN, Via Enrico Fermi, 54 - 00044 Frascati, Italy. \\
 $^3$ EPFL-LPAP, Lausanne, Switzerland.
}%

\date{\today}

\begin{abstract}

The temporal resolution of MeV ultrafast electron diffraction (UED) is fundamentally constrained by the electron bunch length at the sample, motivating the development of new electron sources capable of producing femtosecond-scale bunches. In this work, we propose a multi-cell RF photogun that has a tailored phase velocity profile to generate 5–15~fs~rms MeV electron bunches directly from the electron gun, eliminating the need for downstream compression. This approach achieves comparable performance to conventional one-and-a-half cell photoguns with downstream compression, while reducing system size, complexity, and power requirements.
We examine two implementations: a standing-wave (SW) and a travelling-wave (TW) design. The TW variant demonstrates over an order of magnitude lower power dissipation than typical SW structures, enabling potential kHz operation. When paired with SwissFEL-style C-band RF sources, which offer high amplitude and phase stability, the TW photogun is projected to deliver a temporal resolution of 26~fs~rms.

\end{abstract}

\maketitle

\section{Introduction}
The investigation of atomic and molecular dynamics at ultrafast timescales underpins advances in physical chemistry, materials science, condensed-matter physics, and molecular biology. Free-electron lasers (FELs) have enabled structural studies on femtosecond to picosecond timescales with exceptional spatiotemporal resolution. Their electron-based counterpart, ultrafast electron diffraction (UED), uses femtosecond pump lasers and probe electron bunches to achieve comparable temporal resolution~\cite{Filippetto2022}. Due to their large scattering cross-section, electrons in UED can probe low-Z or thin materials inaccessible to X-rays, though they struggle with thick samples where X-rays remain superior.
\\
Recent efforts in UED have focused on MeV-scale electron beams to extend applicability to thicker samples. The most common sources for such beams are normal-conducting RF photoguns (NC RF photoguns), already implemented in a handful of MeV UED facilities~\cite{Weathersby2015,Hada2011} and proposed for future ones~\cite{Song2022}. These electron guns offer high brightness, but achieving sub-picosecond bunch durations at charges of 100~fC or greater remains challenging due to limited longitudinal compression in these standard configurations.
\\
To achieve shorter bunch lengths at higher bunch charges, MeV UED beamlines have explored RF- and magnetic-based compression schemes. RF compression has proven effective, with demonstrations of 10~fs bunches~\cite{Maxson2017}, but becomes less power efficient in the MeV regime due to the already relativistic velocity of the bunch at the time of compression. 
\\
Magnetic compression is also promising and features in proposed MeV UED facilities~\cite{McKenzie2023}, but it typically involves larger machine footprints. These constraints make both methods less ideal for compact, low-power UED beamlines that would be attractive to small scale facilities.
\\
This raises the question: can a single device provide high-brightness injection (via near on-crest bunch generation), acceleration to MeV energies, and longitudinal compression to femtosecond-scale bunches, while maintaining compactness and low power dissipation? This work addresses this question by introducing a multi-cell RF photogun that leverages a tailored phase velocity profile to generate femtosecond, MeV-scale bunches for UED applications.

\section{Acceleration and Compression via a Tailored Phase Velocity Profile}

We aim to generate high brightness, MeV, femtosecond-scale electron bunches using a single RF device, specifically, a multi-cell RF photogun. This requires tailoring the phase velocity across the device to: accelerate, mitigating space-charge blow-up, and compress the bunches. Such a device integrates high-field photoemission and bunch compression into one structure.
\\
To demonstrate the mechanism for this, we look at the electric field within an travelling-wave structre, written as~\cite{wangler08}:

\begin{equation}
E_z(z) = E_0(z)\cos\left[\int_0^z \left( \frac{\omega}{v_e(z_j)} - \frac{\omega}{v_p(z_j)} \right)dz_j + \phi_0\right].
\label{Eqn:Efield_new}
\end{equation}

where $\omega$ is the angular frequency. Assuming a constant field amplitude ($E_0(z) = E_a$), emission on-crest ($\phi_0 = 0$), and a constant difference between the particle and phase velocity ($v_e-v_p = a$), the kinetic energy (K) gained by a particle in this electric field is found to follow the proportionality:

\begin{equation}
K \propto \int_{z_i}^{z_f} \cos(az)dz.
\end{equation}

Substituting $u = az$, and targeting a phase change (for the centre of the bunch) from $u=0$ to $u=\pi/2$, but with a small phase-offset $d\phi$ representing the finite bunch length, we find:

\begin{equation}
K \propto \int_{0+d\phi}^{\pi/2+d\phi} \cos(u)du \approx 1 - d\phi.
\label{Eqn:Relative_Energy}
\end{equation}

\begin{figure}[!tb]
    \centering
    \includegraphics[width=0.9 \columnwidth]{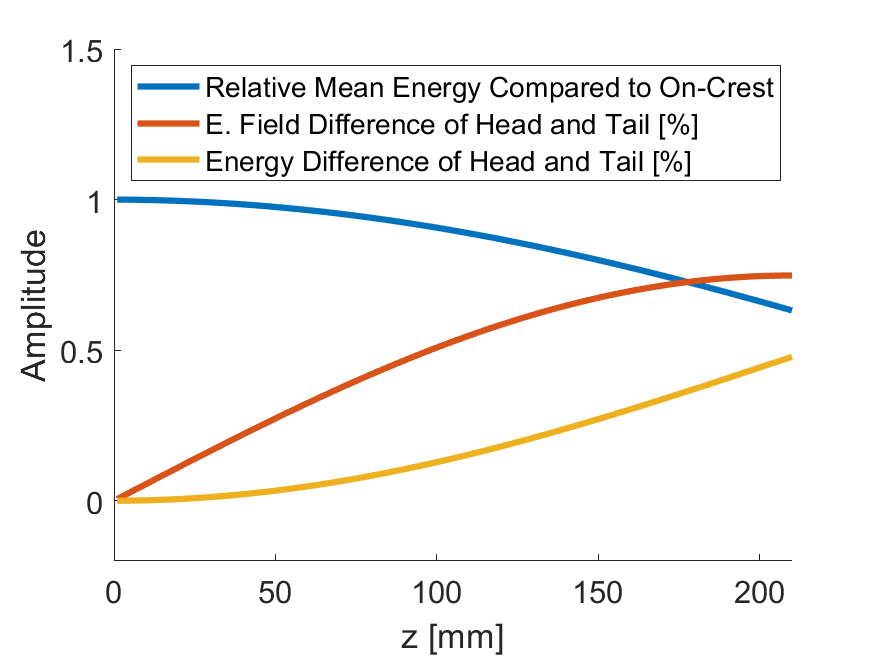}
    \caption{Energy related properties of a 250 fs bunch in a 5.712 GHz electron gun with a tailored phase velocity.}
    \label{Fig:Fig1_1}
\end{figure}

This tells us that using this method, electrons in the tail ($d\phi < 0$) gain more energy than those in the head ($d\phi > 0$), producing an energy chirp and as a consequence the bunch experiences ballistic bunching downstream of the gun. \\ A similar mechanism can be achieved through a variation in the cell length rather than the phase velocity of either standing-wave or travelling-wave systems to induce such a phase slippage, which would replace the integral in Eq.~\ref{Eqn:Efield_new}.\\
\indent Fig.~\ref{Fig:Fig1_1} gives an example where a bunch with an initial bunch length of 250~fs slips forward over the a C-band (5.712 GHz) RF phase to induce this energy chirp, and therefore leads to compression. We see that after 210~mm the relative energy spread between the head and the tail is 0.5$\%$.

\section{Phase Velocity Profile for Femtosecond Bunches}
\label{Subsec:multicell_rf_gun}
\begin{figure}[!tb]
    \centering
    \begin{subfigure}{0.45\textwidth}
    \includegraphics[width=\textwidth]{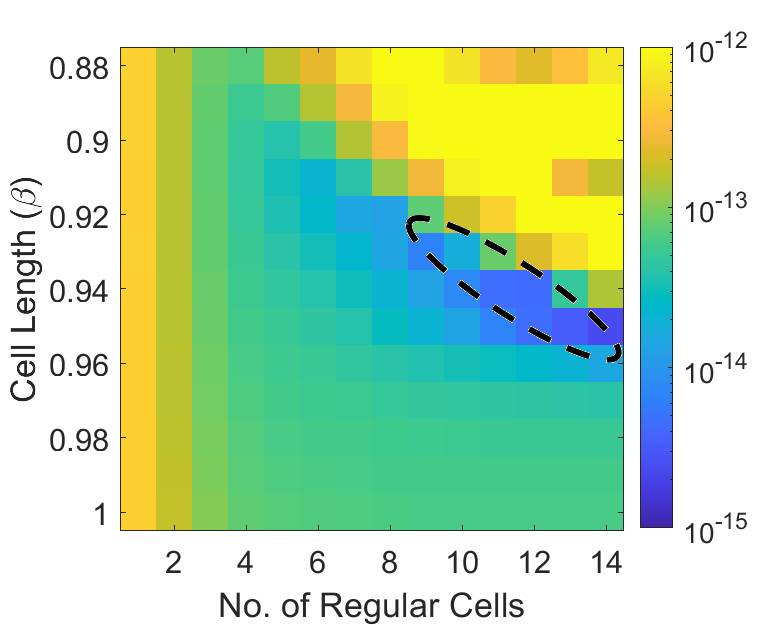}
    \caption{Bunch Length in seconds}
    \label{Subfig:Fig2a}
    \end{subfigure}
     \begin{subfigure}{0.45\textwidth}
    \includegraphics[width=\textwidth]{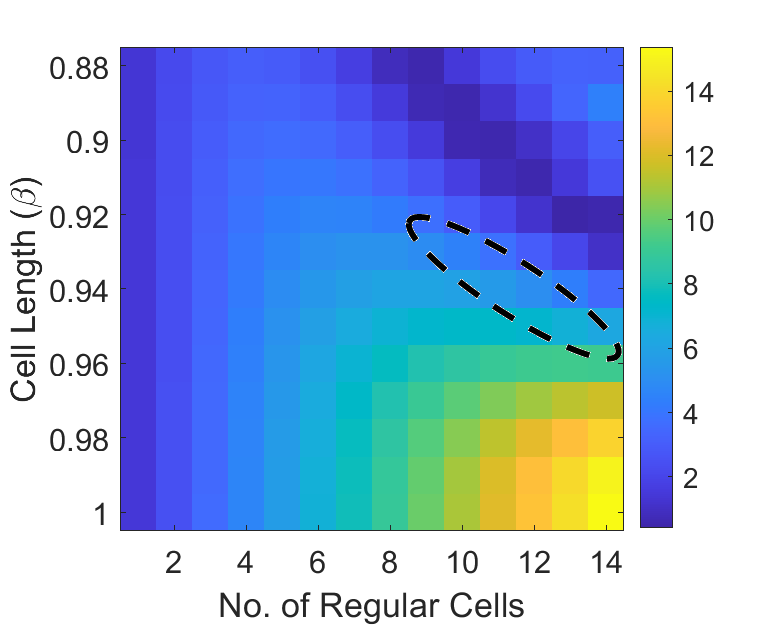}
    \caption{Final Energy in MeV}
    \label{Subfig:Fig2b}
    \end{subfigure}
    \caption{Scan of the number of regular cells after the first half cell (x-axis) and the cell-length reduction (y-axis), used in the cell configuration investigation. The modified cell length is expressed as a ratio of the synchronous velocity to the speed of light, expressed as the relativistic beta ($\beta$).}
    \label{Fig:Fig2}
\end{figure}

To validate the beam dynamics and identify the optimal phase velocity/cell-length to guide the RF photogun design, particle tracking simulations were performed using $N$ single cells concatenated into a full RF photogun field map~\cite{Schaer2023}, with the cells modeled in CST~\cite{CST}. The first cell was set as a half-cell for cathode placement. The tailored phase velocity difference was introduced by varying cell-lengths in MATLAB using the method describe in~\cite{Schaer2023} and, as described above, this is the equivalent of a phase velocity change. \\
Optimisation focused on minimising the bunch length 1–3~m downstream of the cathode ($z=0$). Figure~\ref{Fig:Fig2} shows the minimum bunch length and mean bunch energy as functions of the number of regular cells (1–14, excluding the first half-cell) and the normalised cell length (1 to 0.88, the latter of which corresponds to a 12\% reduction in length). When using 9–14 regular cells with a cell-length reduction of 5–8\%, the bunch length drops below 10~fs (circled region in Fig.~\ref{Subfig:Fig2a}, noting the logarithmic color scale). Beyond this “optimal compression zone” (top-right of Fig.~\ref{Subfig:Fig2a}), too much compression causes the minimum bunch length to shift upstream of the target 1–3~m region, seen as a longer bunch length in the region of interest.

In this strong compression region (also circled in Fig.~\ref{Subfig:Fig2b}), the mean energy reduces to 3–6~MeV due to the trade-off between acceleration and compression, which aligns with desirable energy for MeV-UED~\cite{Filippetto2022}. The optimal configuration was found with 11.5 cells and a normalised cell-length of 0.94 (recalling this is equivalent to changing the phase velocity to 0.94c). A solenoid placed 0.3~m downstream of the cathode mitigated transverse space-charge effects. All simulations included space-charge effects, used 5000 macro-particles, and were carried out with GPT~\cite{gpt_cite}. For this step, the initial bunch parameters were based on the SLAC MeV UED beamline~\cite{Weathersby2015} (see their Table~\ref{Tab:Tab1}) without further optimisation.

\begin{figure*}[!tb]
    \centering
    \begin{subfigure}[b]{0.45\textwidth}
         \centering
             \includegraphics[width= \linewidth]{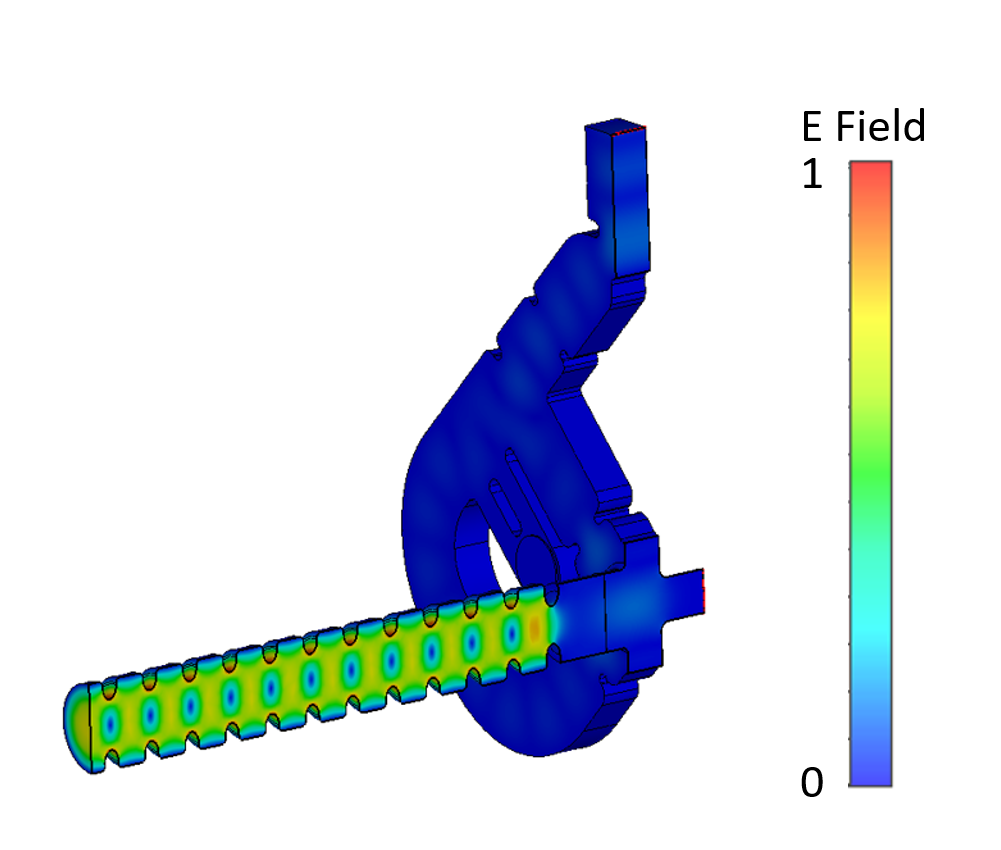}
         \caption{Standing-wave electric field magnitude.}
         \label{Subfig:Fig3a}
     \end{subfigure}
     \begin{subfigure}[b]{0.45\textwidth}
         \centering
             \includegraphics[width= \linewidth]{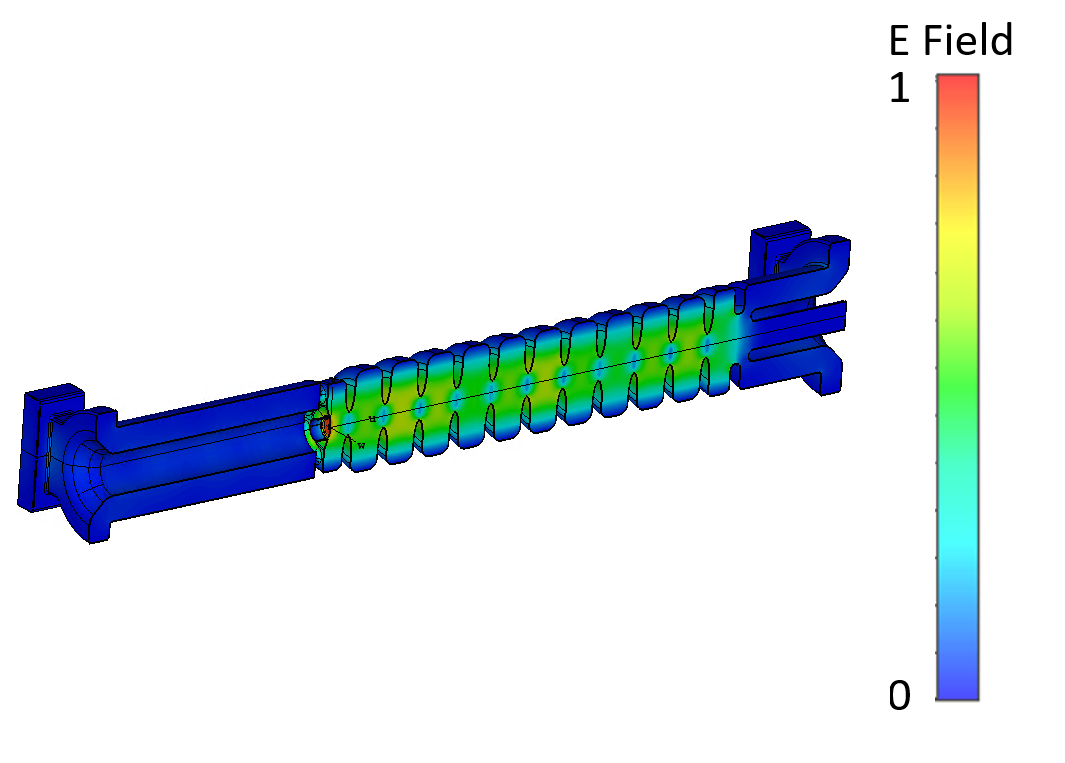}
         \caption{Travelling-wave electric field magnitude.}
         \label{Subfig:Fig3b}
     \end{subfigure}   
     \begin{subfigure}[b]{0.45\textwidth}
         \centering
             \includegraphics[width= \linewidth]{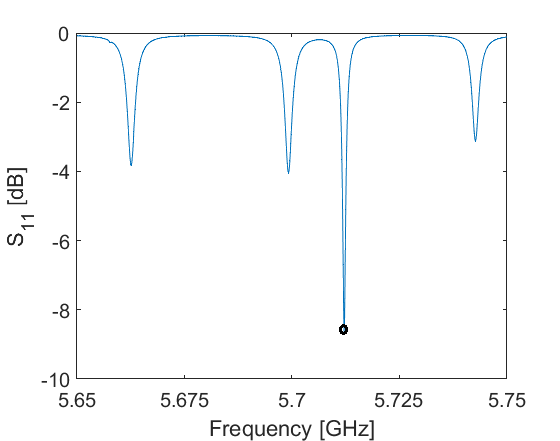}
         \caption{Standing-wave scattering parameters.}
         \label{Subfig:Fig3c}
     \end{subfigure}   
     \begin{subfigure}[b]{0.45\textwidth}
         \centering
             \includegraphics[width= \linewidth]{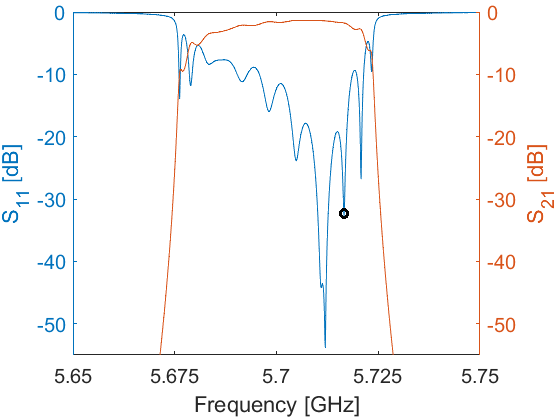}
         \caption{Travelling-wave scattering parameters.}
         \label{Subfig:Fig3d}
     \end{subfigure}   

    \caption{Design of the standing-wave (left) and travelling wave (right) RF photoguns illustrating the electric field in the vacuum model (top row) and the scattering (S-) parameters (bottom row)~\cite{Lucas2023}. The operating points on the S-parameter diagrams are marked with a black circle. The TW gun operates away from its nominal 5.712 GHz to change the phase velocity.}
    \label{Fig:Fig6}
\end{figure*}

\section{Electron Sources with Tailored Phase Velocity}
Given the two approaches towards achieving phase-particle velocity difference, that being a change to the phase velocity or cell length, here we propose two separate gun designs each exploiting the methodologies, separately. Based on the standing-wave RF photogun of~\cite{Giribono2023}, the first proposed electron gun design, illustrated in Fig.~\ref{Subfig:Fig3a}, is a multi-cell C-band standing-wave RF photogun. It has 11.5-cells designed to operate at a peak cathode field of 80~MV/m. The cells are designed with very large iris radii, compared to the original 2.5-cell SW gun, in order to increase the mode separation~\cite{Limborg2005}. With this new design, the mode separation between the $\pi$-mode and its closest neighbouring mode was increased to 12~MHz (Fig.~\ref{Subfig:Fig3c}). The mode launcher features four input ports, as was used in the original design from~\cite{Giribono2023}. To tailor the phase velocity, the cell-lengths are reduced by 6$\%$, determined in the previous section to be a suitable cell-length reduction, from the original lengths.\\ \indent
The second proposed RF design is the novel RF design of the C-band travelling-wave RF photogun. This gun has been realised as part of the IFAST project~\cite{ifast} (Fig.~\ref{Subfig:Fig3b}). A comprehensive report on its RF design is given in~\cite{Lucas2023} therefore here we simply report on the unique operation to achieve the tailored phase velocity. Rather than usual operation, the TW photogun is driven at a frequency 3~MHz greater than its original operating frequency. Additionally, it operates at a temperature 10$^\circ C$ above the original design temperature, reducing the nominal resonant frequency by 1 MHz. Consequently, the equivalent driver frequency is 5.716~GHz, where the TW gun maintains a reflection coefficient ($S_{11}$) below -30 dB (See Fig.~\ref{Subfig:Fig3d}). Such operation reduces the phase velocity appropriately and therefore requires no physical modification to the original design's cell length. Table~\ref{Tab:Tab1} lists the RF parameters of the TW and SW guns, for the tailored phase velocity. 

\begin{table}[htb!]
\caption{RF parameters of the 11.5 cell SW and TW RF photoguns. The parameters with bold text are those that relate to the phase velocity/cell-length difference. The TW gun is separated into low (LP) and high power operation (HP).}
\centering
\resizebox{\columnwidth}{!}{%
\begin{tabular}{||c | c c c||} 
\hline
Parameter & TW Gun & SW Gun & Units \\ [0.5ex] 
\hline\hline
 Mode Frequency & \textbf{5.711} & 5.712 & GHz \\
 RF Driver Frequency & \textbf{5.715} & 5.712 & GHz \\
 No. of Accelerating Cells & 11.5 & 11.5 &  \\
 Cell-length & 17.495 & \textbf{16.445} & mm  \\
 Structure Active Length & 250 & 380 &mm  \\
 \hline
 Peak Power & 10 (LP)/20(HP) & 23 & MW \\
 Peak electric field at cathode & 80/113 & 80 & MV/m \\
 Filling-Time & 90 & 448 & ns \\
 RF Pulse Length & 90 & 1344 & ns \\
 Mean RF Power (100 Hz) & 90/180 & 3100 & W \\
 Power Dissipated in Gun (100 Hz) & 29/58 & 2554 & W \\
\hline
\end{tabular}
}
\label{Tab:Tab1}
\end{table}

\begin{figure}[!tb]
    \centering
    \includegraphics[width= \linewidth]{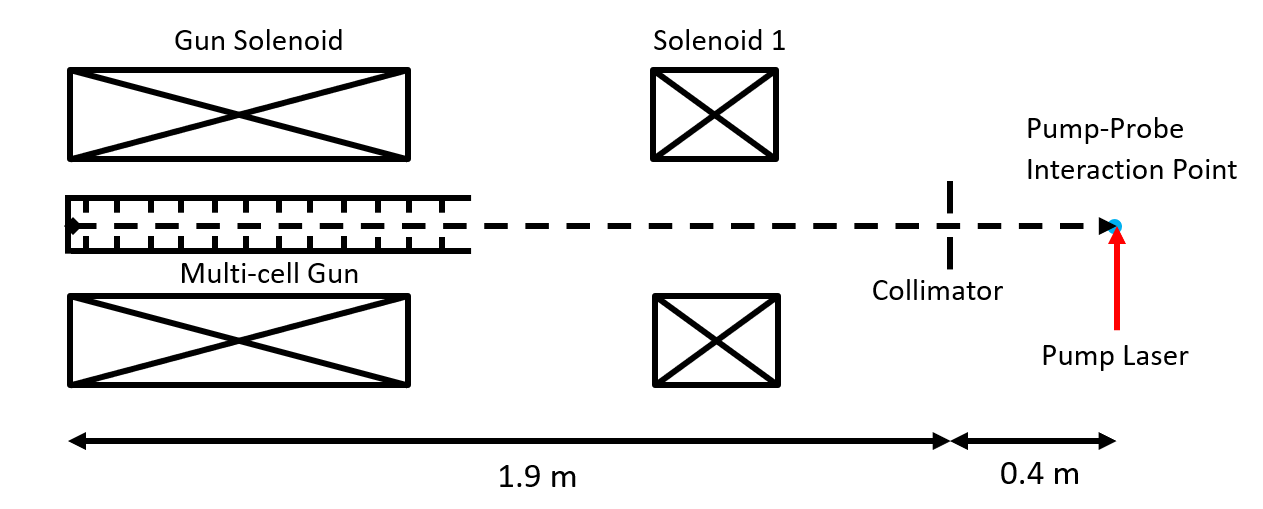}
    \caption{Beamline layout for the particle tracking simulations consisting of the multi-cell RF photogun with its solenoid, along with a solenoids downstream to focus the beam onto the sample.}
    \label{Fig:Fig17b}
\end{figure}

\begin{figure*}[!tb]
    \centering
     \begin{subfigure}[b]{0.45\textwidth}
         \centering
             \includegraphics[width= \linewidth]{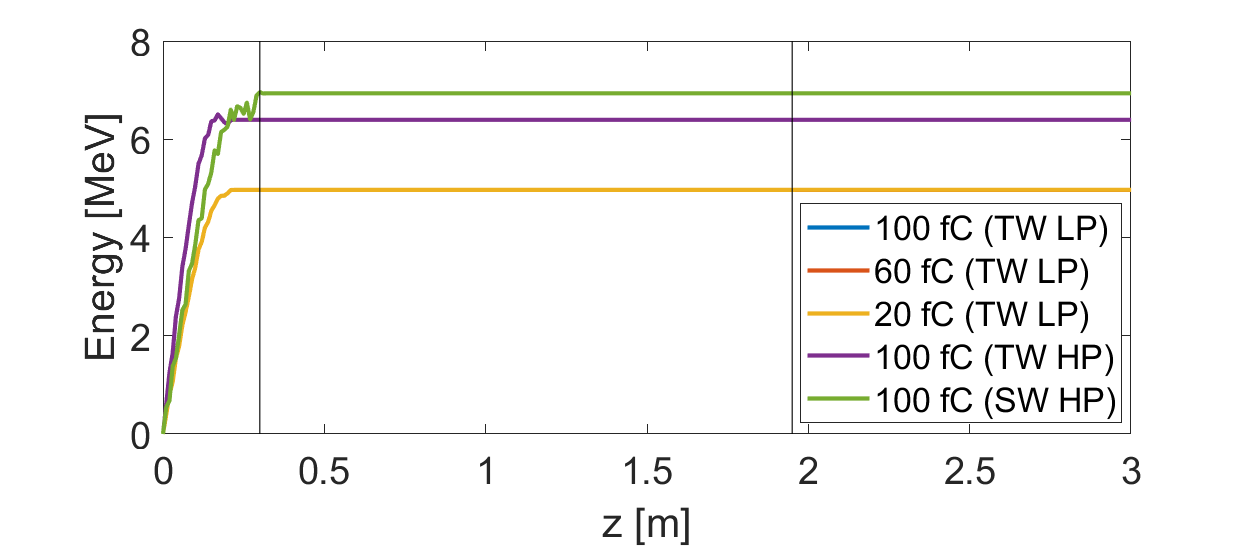}
         \caption{Mean Energy}
         \label{Subfig:Fig13a}
     \end{subfigure}   
     \begin{subfigure}[b]{0.45\textwidth}
         \centering
             \includegraphics[width= \linewidth]{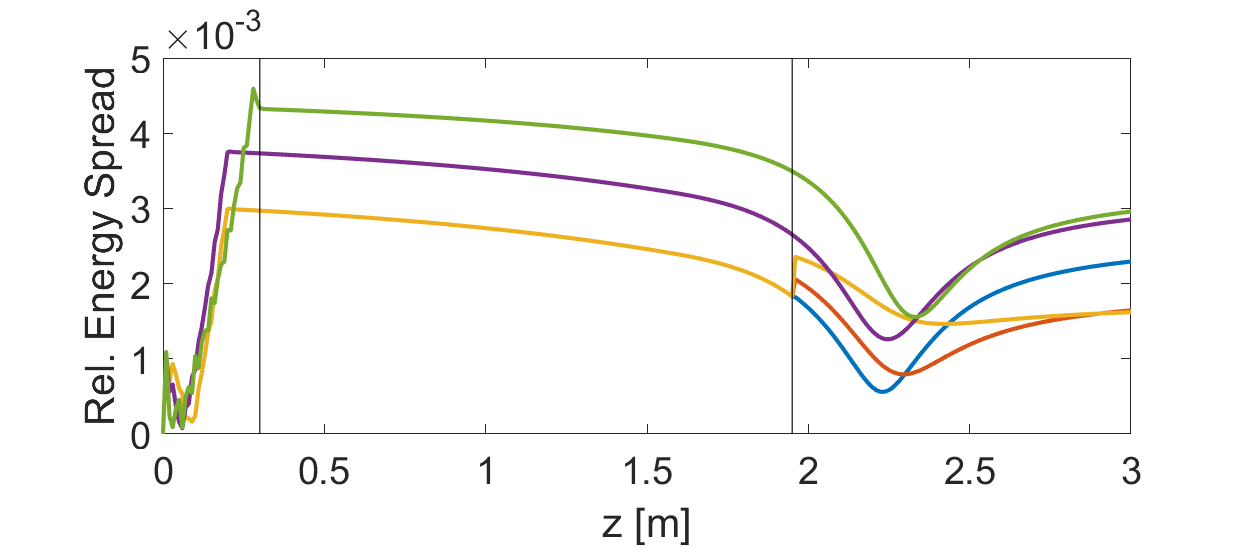}
         \caption{Energy Spread}
         \label{Subfig:Fig13b}
     \end{subfigure} 
     \begin{subfigure}[b]{0.45\textwidth}
         \centering
             \includegraphics[width= \linewidth]{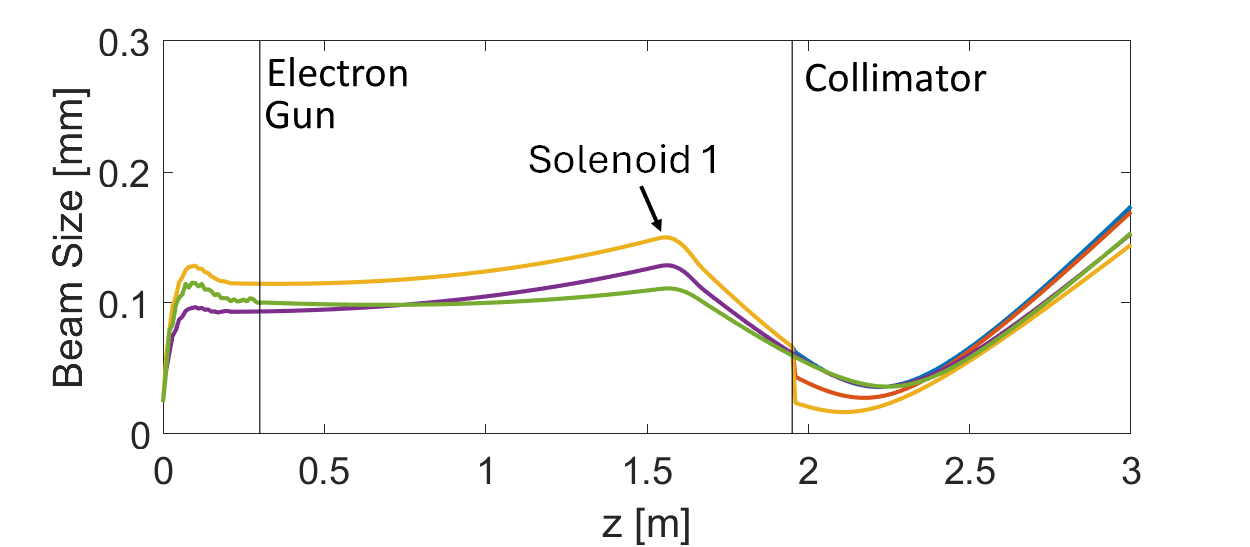}
         \caption{Bunch Size}
         \label{Subfig:Fig13c}
     \end{subfigure}   
     \begin{subfigure}[b]{0.45\textwidth}
         \centering
             \includegraphics[width= \linewidth]{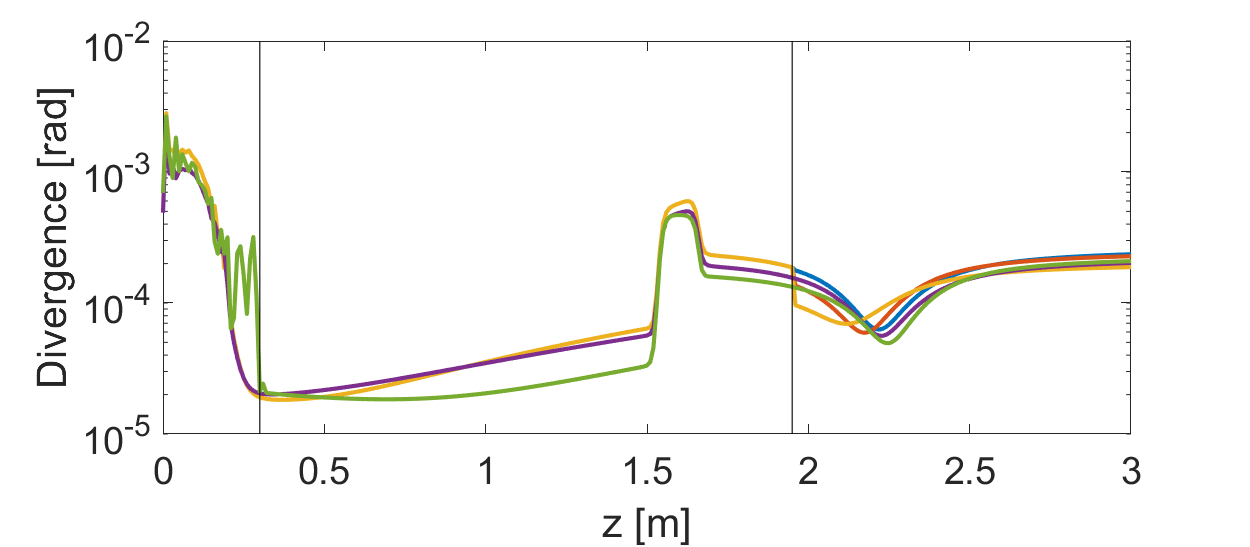}
         \caption{Divergence}
         \label{Subfig:Fig13d}
     \end{subfigure}
     \begin{subfigure}[b]{0.45\textwidth}
         \centering
             \includegraphics[width= \linewidth]{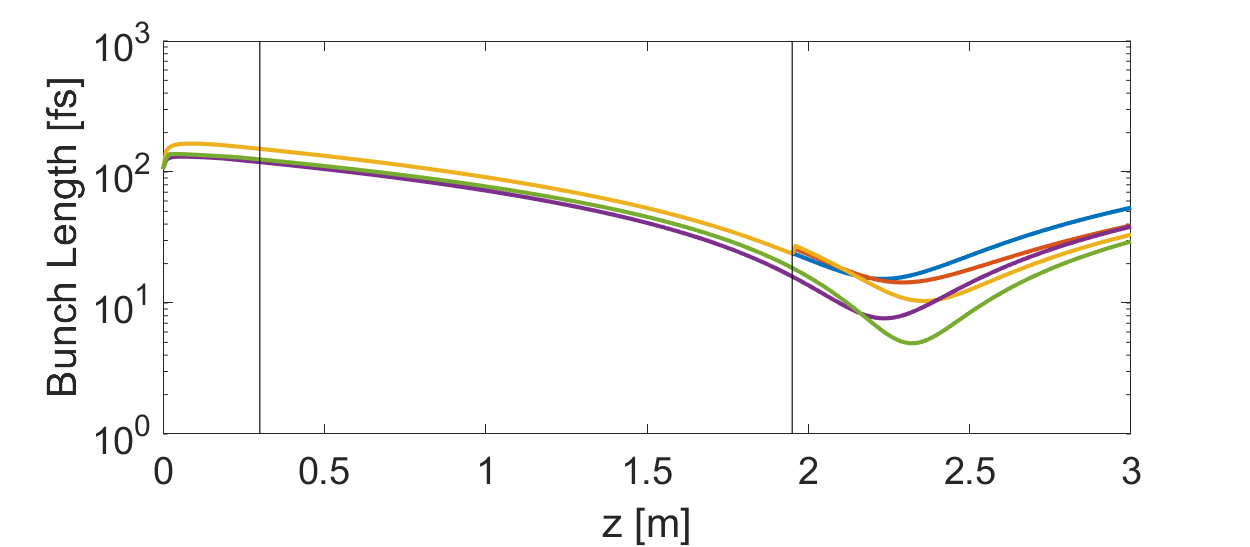}
         \caption{Bunch Length}
         \label{Subfig:Fig13e}
     \end{subfigure}   
     \begin{subfigure}[b]{0.45\textwidth}
         \centering
             \includegraphics[width= \linewidth]{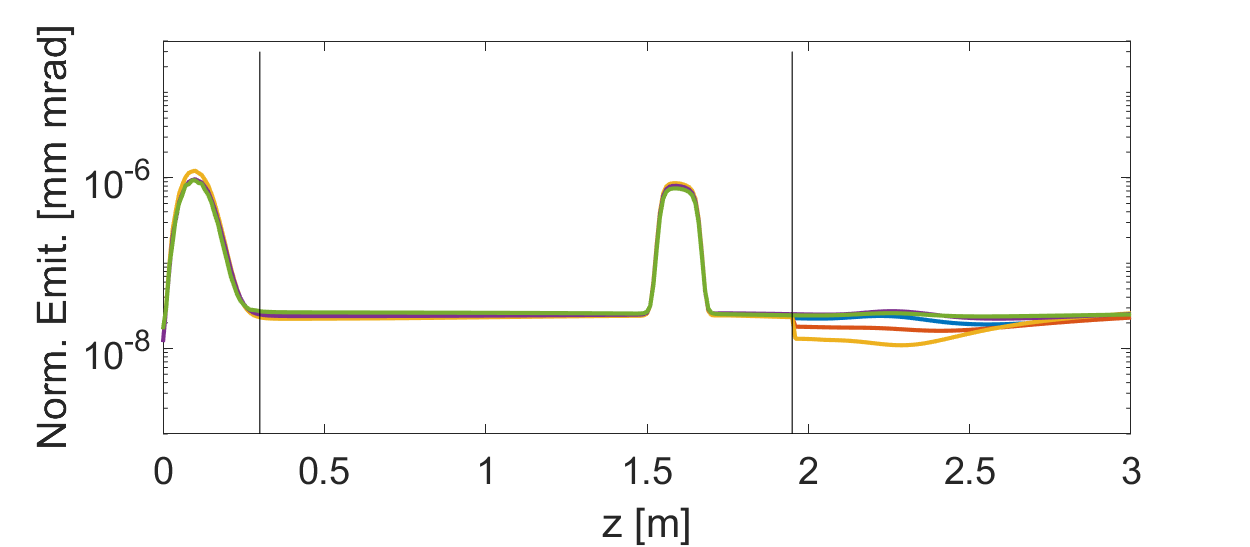}
         \caption{Normalised Emittance}
         \label{Subfig:Fig13f}
     \end{subfigure} 
    \caption{Bunch parameters (rms) plotted against the longitudinal position in the beamline. The location of the end of the electron gun and the collimators are indicated with verical lines while the final focus solenoid's position is annotated on the beam size plot.}
    \label{Fig:Fig13}
\end{figure*}

\begin{table*}[!htb]
 \caption{Summary of the properties at photo-cathode laser driving the multi-cell RF photoguns and the resultant beam quality at the interaction point. The terms `TW' for travelling-wave, `SW' for standing-wave, `LP' and `HP' for low and high power, respectively.}
\centering
 \begin{tabular}{||c | c  c  c  | c | c ||} 
 \hline
 Parameter  &  & TW LP & & TW HP & SW HP \\ [0.5ex] 
 \hline\hline
 Peak electric field at cathode [MV/m] &  & 80 &  & 113 & 80 \\
 Electric field at cathode at extraction RF phase [MV/m] &  & 50 &  & 84 & 51.5 \\
 Cathode Bunch Charge [fC] & 100 & 100 & 100 & 100 &100\\
 Transverse Laser profile &  & Gaussian &   & Gaussian & Gaussian \\
 Longitudinal Laser profile &  & Gaussian &   & Gaussian & Gaussian \\
 Laser Beam Radius (RMS) [$\mu$m] &  & 25 &  & 25 & 25 \\
 Intrinsic Thermal Emittance [mm mrad/mm] &  & 0.55 &  & 0.55 & 0.55\\
 Laser Pulse Length (FWHM) [fs] &  & 250 &  & 250 & 250 \\
 \hline
 Charge at IP [fC] & 100 & 60 & 20 & 100 & 100 \\
 Collimator Diameter [\textmu m] & 400 & 200 & 100 & 400 & 400 \\
 Mean Energy [MeV] & 4.96 & 4.87 & 5.01 & 6.4 & 6.93\\
 Relative Energy Spread [\%] & 0.05 & 0.07 & 0.14 & 0.12 & 0.15\\
 Bunch size (RMS) [$\mu$m] & 36 & 27 & 17 & 36 & 36 \\
 Bunch length (RMS) [fs] & 15 & 14 & 10 & 8 & 4.9 \\
 Normalised Emittance (RMS) [nm mrad] & 24 & 17 & 11 & 27 & 26 \\
 Divergence (RMS) [$\mu$rad] & 63 & 59 & 70 & 56 & 49\\
 \hline
 \end{tabular}
 \label{Tab:Tab2}
\end{table*}

\section{MeV UED Beamline Performance}
\label{Sec:BeamlinePerformance}

Using the field maps of the two RF designs and the GPT code, we assess the performance of both photoguns as electron sources for a MeV UED beamline. The beam is tracked from the cathode through to and beyond the interaction point, with one-to-one modelling of the electrons, i.e. not using macroparticles.
\\
The proposed beamline layout is shown in Fig.~\ref{Fig:Fig17b}. The gun solenoid design is that used in~\cite{Lucas2023}, while solenoid 1 is modeled analytically. Simulation parameters for the RF photoguns and laser are listed in Tables~\ref{Tab:Tab1} and~\ref{Tab:Tab2}, respectively. Figure~\ref{Fig:Fig13} shows the evolution of rms bunch properties across five scenarios: three with the TW photogun and a collimator at 1.9~m adjusting the delivered bunch charge (100~fC, 60~fC, and 20~fC); one TW case with increased input power; and one with the SW photogun for comparison and to demonstrate the equivalence of the cell length versus phase velocity methodologies.
\\
The interaction point is defined where both transverse and longitudinal beam sizes are minimised, optimising sample overlap and temporal resolution. Table~\ref{Tab:Tab2} summarises beam properties at this point. For the low-power TW case, 5~MeV, 100~fC bunches with 0.1\% rms energy spread, 15~fs rms bunch length, and 24~nm rms emittance are illustrated. Increasing the input power reduces the bunch length to 8~fs. The energy spread from the TW gun at 0.3 m (0.3–0.45\% rms) aligns well with the predictions from the analytical model (0.5\% rms) demonstrated in Fig.~\ref{Fig:Fig1_1}, validating the theoretical framework.
\\
In the SW case, the mean bunch energy is slightly higher due to a slightly larger integrated electric field of the device, but beam performance is comparable to the TW design, demonstrating the equivalence of both approaches to inducing an energy chirp using a tailored phase velocity.

\section{Temporal Resolution}
The temporal resolution ($\Delta t_{\text{res}}$) of a UED experiment defines the shortest observable event duration and depends on both the probe electron bunch and the pump laser. It is given by the quadrature sum of uncorrelated contributions~\cite{Filippetto2022}:

\begin{equation}
    \Delta t_{\text{res}} = \sqrt{\Delta t_{\text{electron}}^2 + \Delta t_{\text{pump}}^2 + \Delta t_{\text{vm}}^2 + \Delta t_{\text{jitter}}^2},
    \label{Eqn:resolution}
\end{equation}

where $\Delta t_{\text{electron}}$ is the rms electron bunch length at the interaction point, $\Delta t_{\text{pump}}$ the rms pump laser pulse duration, $\Delta t_{\text{vm}}$ the velocity-mismatch contribution, and $\Delta t_{\text{jitter}}$ the arrival-time jitter between pump and probe. All values are rms.
\\
In this MeV-UED beamline, the target is a temporal resolution toward the femtosecond regime. The electron bunch and pump laser pulse durations are each 15~fs~rms. Due to the MeV-scale electron energy, the velocity mismatch contribution is negligible (a few femtoseconds for a 100~\textmu m sample).
\\
The final source uncertainty arises from $\Delta t_{\text{jitter}}$. Although the photocathode and pump lasers can be derived from a common source, RF amplitude and phase jitter, primarily from the amplifiers, introduce arrival-time jitter. Based on measured C-band RF station stabilities at SwissFEL~\cite{Geng2022,Beard2022,Lucas2025}, the relative amplitude and absolute phase stabilities have median values of $7.5 \times 10^{-5}$ and 0.026$^\circ$ at 5.712~GHz, respectively.
\\
To estimate $\Delta t_{\text{jitter}}$, 500 GPT simulations were performed with Gaussian-distributed amplitude and phase perturbations applied to the TW photogun field map. The resulting pump-probe arrival-time jitter was 16~fs~rms, dominated by the phase noise; the amplitude stability contributed only 1~fs~rms. Combining all terms in Eqn.~\ref{Eqn:resolution}, the predicted temporal resolution is 26~fs~rms.

\section{Comparison between Standing-Wave vs Travelling-Wave}
Both standing-wave (SW) and travelling-wave (TW) RF photoguns can realise a tailored phase-particle velocity difference. However, their operational characteristics differ significantly. The TW gun benefits from a short filling time and low attenuation, resulting in low power dissipation within the device. With a 90~ns filling time, it can operate at 80~MV/m cathode gradient using a 90~ns RF pulse and 10~MW input power. At 100~Hz, this corresponds to a mean input power of 90~W, with only 29~W dissipated as resistive losses in the gun walls. The remaining power is absorbed by a downstream RF load, which is less sensitive to temperature. This low power dissipation means that, with the right power source, this device could open up operation into the kHz regime.
\\
In contrast, the SW photogun requires 23~MW input power to achieve a similar gradient, and has a much longer filling time of 448~ns. As steady-state typically requires $\sim$3 filling times (see Table~\ref{Tab:Tab1}), and with an $S_{11}$ of -8~dB, the power dissipated in the cavity walls reaches $\sim$2.5~kW—nearly two orders of magnitude higher than in the TW case.
\\
Additionally, the SW gun's longer RF pulse duration limits its peak electric field capabilities, due to the inverse correlation between breakdown threshold and pulse length~\cite{Grudiev2009}. This implies the TW gun may support higher cathode gradients or offer improved breakdown reliability at the same field level.
\\
Another key distinction lies in tunability. Once constructed, the SW gun’s longitudinal focal point can only be adjusted by changing the cathode phase or input power, both of which impact field conditions and require re-optimisation. The TW gun, however, allows more flexible tuning via changes in drive frequency or structure temperature, which modify the phase velocity and hence the phase-particle velocity difference.
\\
Finally, the standing-wave gun requires a high-power RF circulator to protect the power source from reflections. However, such devices at C-band are difficult to fabricate and not commercially available at the necessary power levels.

\section{Conclusions}
A multi-cell RF photogun that uses tailored phase-particle velocity difference to generate femtosecond electron bunches has been proposed as an electron source for an MeV UED beamline. Compared to its one-and-a-half-cell counterparts, such an electron source can achieve significantly better longitudinal compression allowing for the gun to produce electron bunches down to the 5-15 fs regime for bunch charges of 100~fC and a low transverse emittance of 24~nm~rad. This is comparable to using a more conventional one-and-a-half cell RF photogun with an additional bunch compression scheme downstream. Two designs for such an electron gun were proposed with each illustrating similarly high performance. The TW version requires only 90~ns, 10~MW RF pulses to drive the system resulting just 29 W of power dissipated in the gun. This allows for more energy-efficient operation and also opens up the possibility of operating at repetition rates significantly higher than SW RF photoguns.

\section{Acknowledgements}
The authors thank Xijang Wang for his comments and constructive conversations on MeV-UED.

\section{Funding}
This project has received funding from the European Union’s Horizon 2020 Research and Innovation program under GA No101004730.

\section{Appendix}
Here we include the full derivation of Eqn.~\ref{Eqn:Relative_Energy}. The general equation of the travelling-wave seen by an electron in a waveguide is written in a general form as~\cite{wangler08}

\begin{equation}
    E_z(z,t) = E_0(z)\cos \bigg [\omega t-\int_0^z dz_0 k(z_0) +\phi_0 \bigg],
    \label{Eqn:Efield}
\end{equation}

where $E_0(z)$ is the longitudinal electric field amplitude profile, $\omega$ is the angular frequency, $k(z)$ is the wavenumber and $\phi_0$ is the initial phase of the electron on the RF. We can assign the time parameter ($t$) with respect to the longitudinal position ($z$) of an electron travelling with velocity $v_e(z)$ as:

\begin{equation}
    t(z) = \int_0^z dz_i \frac{1}{v_e(z_i)}.
\end{equation}

Similarly, the wavenumber of the RF can be written in terms of its phase velocity, $v_p(z)$, such that:

\begin{equation}
    k(z) = \frac{\omega}{v_p(z)}.
\end{equation}

With these two relations, we can rewrite Eqn.~\ref{Eqn:Efield} as:

\begin{equation}
    E_z(z) = E_0(z)\cos \bigg [\int_0^z \bigg ( \frac{ \omega }{v_e(z_j)}- \frac{\omega}{v_{p}(z_j)} \bigg ) dz_j +\phi_0 \bigg].
\end{equation}

With this (Eqn.~\ref{Eqn:Efield_new} from main text) can therefore be written more simply as:

\begin{equation}
    E_z(z) = E_0(z)\cos (\alpha (z) +\phi_0 ).
    \label{Eqn:Efield_new2}
\end{equation}

We can now look at the energy imparted on the electron. We can express this energy (K) as
\begin{equation}
    K = q\int_{z_i}^{z_f} E_z(z) dz
\end{equation}
where $q$ is the particle charge and $\gamma$ is the lorentz factor. For $E_z(z)$ we can substitute in Eqn.~\ref{Eqn:Efield_new2}, assume $E_0(z) = E_a$, $\phi_0=0$, $\alpha(z) = az$. This gives:
\begin{equation}
    K \propto\int_{z_i}^{z_f} cos(az) dz
\end{equation}
Integrate over a phase advance from on-crest to zero-crossing:
\begin{equation}
    K \propto \int_{0+d\phi}^{\pi/2+d\phi} \cos(u) du \approx 1-d\phi
\end{equation}
where we assume $d\phi$ is small but non-zero.
\\

\indent For a set of SW cavities, achieving this phase shift per cell requires changing the lengths of the individual cells to modify the time-of-flight of the bunch within each cell~\cite{Vretenar2012}. The electric field at the nth cells, as seen by the electron, in the case where each cell induces an arbitrary phase shift ($\Delta\phi_n$) can be written as:
\begin{equation}
    E_{z,n}(z) = E_{0,n}(z)\cos (\sum_{m=1}^n \Delta\phi_m + \phi_0).
\end{equation}
This is the analogue of Eqn.~\ref{Eqn:Efield_new2} using cell-length modification which can be achieved in both  standing-wave and travelling-wave devices.


\begin{thebibliography}{9}

\bibitem{Filippetto2022}
Filippetto, D. and Musumeci, P. and Li, R. K. and Siwick, B. J. and Otto, M. R. and Centurion, M. and Nunes, J. P. F., Ultrafast electron diffraction: Visualizing dynamic states of matter, Rev. Mod. Phys. 94.045004 (2022) 10.1103/RevModPhys.94.045004.

\bibitem{Weathersby2015}
S. P. Weathersby, G. Brown, M. Centurion, T. F. Chase, R. Coffee, J. Corbett, J. P. Eichner, J. C. Frisch, A. R. Fry, M. Gühr, N. Hartmann, C. Hast, R. Hettel, R. K. Jobe, E. N. Jongewaard, J. R. Lewandowski, R. K. Li, A. M. Lindenberg, I. Makasyuk, J. E. May, D. McCormick, M. N. Nguyen, A. H. Reid, X. Shen, K. Sokolowski-Tinten, T. Vecchione, S. L. Vetter, J. Wu, J. Yang, H. A. Dürr, X. J. Wang; Mega-electron-volt ultrafast electron diffraction at SLAC National Accelerator Laboratory. Rev. Sci. Instrum. 1 July 2015; 86 (7): 073702. https://doi.org/10.1063/1.4926994

\bibitem{Hada2011}
M. Hada, et al, "REGAE: New Source for Atomically Resolved Dynamics," in Research in Optical Sciences, OSA Technical Digest (Optica Publishing Group, 2012), paper JT2A.47.

\bibitem{Song2022}
Development of a 1.4-cell RF photocathode gun for single-shot MeV ultrafast electron diffraction devices with femtosecond resolution.

\bibitem{Maxson2017}
Maxson, Jared and Cesar, David and Calmasini, Giacomo and Ody, Alexander and Musumeci, Pietro and Alesini, David, Direct Measurement of Sub-10 fs Relativistic Electron Beams with Ultralow Emittance, Phys. Rev. Lett. 118, 154802  (2017) 10.1103/PhysRevLett.118.154802.

\bibitem{McKenzie2023}
J. W. McKenzie, et al. Establishing a relativistic ultrafast electron diffraction and imaging (RUEDI) UK National Facility, Proceedings of the International Particle Accelerator Conference, 2023, p. TUPL144.

\bibitem{wangler08}
T. Wangler, RF Linear Accelerators, ISBN: 9783527406807, Wiley (2008).

\bibitem{Vretenar2012}
M. Vretenar, Low-beta structures (2012) \url{https://cds.cern.ch/record/1415909/files/p319.pdf }

\bibitem{Schaer2023}
M. Schaer, et al, rf traveling-wave electron gun for photoinjectors. Phys. Rev. Accel. Beams 19, 072001 https://doi.org/10.1103/PhysRevAccelBeams.19.072001

\bibitem{gpt_cite}
Pulsar Physics General Particle Tracer (GPT)  (2021,1,25), www.pulsar.nl/gpt

\bibitem{CST}
3DS CST Microwave Studio., https://www.3ds.com/

\bibitem{Lucas2023}
T. G. Lucas, et al, Toward a brightness upgrade to the SwissFEL: A high gradient traveling-wave rf photogun. Phys. Rev. Accel. Beams 26, 103401 10.1103/PhysRevAccelBeams.26.103401

\bibitem{Giribono2023}
A Giribono, et al., Dynamics studies of high brightness electron beams in a normal conducting, high repetition rate C-band injector. Phys. Rev. Accel. Beams  26, 083402 https://doi.org/10.1103/PhysRevAccelBeams.26.083402

\bibitem{Limborg2005}
Limborg-Deprey, C., Xiao, L., Dowell, D. H., Li, Z., and Schmerge, J. (2005, May). Modifications on RF Components in the LCLS Injector. In Proceedings of the 2005 Particle Accelerator Conference (pp. 2233-2235). IEEE.

\bibitem{ifast}
Innovation Fostering in Accelerator Science and Technology (I.FAST) https://ifast-project.eu/index.php/

\bibitem{Beard2022}
C. D. Beard, et al., RF SYSTEM PERFORMANCE IN SwissFEL, in: Proceedings of the International Linear Accelerator Conference, 2022, p. TH2AA02

\bibitem{Geng2022}
Z. Geng, et al., RF jitter and electron beam stability in the SwissFEL linac, in: Proceedings of the Free Electron Laser Conference, 2022, p. TH2AA02

\bibitem{Lucas2025}
T.G. Lucas, et al., Multi-year Longitudinal Analysis of Vacuum Breakdown and RF Conditioning in the C-band Linac of SwissFEL, IEEE Transactions On Nuclear Science, February 2024.

\bibitem{Grudiev2009}
A. Grudiev, S. Calatroni and W. Wuensch, New local field quantity describing the high gradient limit of accelerating structures, {\em Phys. Rev. ST Accel. Beams} \textbf{12}, 102001 (2009,10)

\end{thebibliography}
\end{document}